\documentclass[aps,prl,preprint,tightenlines,superscriptaddress,showpacs,byrevtex]{revtex4}
\usepackage{graphicx} 
\usepackage{dcolumn}  

\graphicspath{{ps}}

\renewcommand{\arraystretch}{1.1}

\newcommand{\bzbar}{\ensuremath{\overline{B}{}^0}}
\newcommand{\bbar}{\ensuremath{\overline{B}}}

\newcommand{\kzbarstar}{\ensuremath{\overline{K}{}^{*0}}}

\newcommand{\mev}{\ensuremath{\mathrm{MeV}}}
\newcommand{\gev}{\ensuremath{\mathrm{GeV}}}

\begin{document}


\preprint{\vbox{ \hbox{   }
                 \hbox{BELLE-CONF-0511}
                 \hbox{LP2005-146}
                 \hbox{EPS05-483} 
}}

\title{ \quad\\[0.5cm]  Improved Measurements of 
{\boldmath $\bzbar$} $\rightarrow$ {\boldmath $D_{\it sJ}^+ K$}$^-$ decays}

\affiliation{Aomori University, Aomori}
\affiliation{Budker Institute of Nuclear Physics, Novosibirsk}
\affiliation{Chiba University, Chiba}
\affiliation{Chonnam National University, Kwangju}
\affiliation{University of Cincinnati, Cincinnati, Ohio 45221}
\affiliation{University of Frankfurt, Frankfurt}
\affiliation{Gyeongsang National University, Chinju}
\affiliation{University of Hawaii, Honolulu, Hawaii 96822}
\affiliation{High Energy Accelerator Research Organization (KEK), Tsukuba}
\affiliation{Hiroshima Institute of Technology, Hiroshima}
\affiliation{Institute of High Energy Physics, Chinese Academy of Sciences, Beijing}
\affiliation{Institute of High Energy Physics, Vienna}
\affiliation{Institute for Theoretical and Experimental Physics, Moscow}
\affiliation{J. Stefan Institute, Ljubljana}
\affiliation{Kanagawa University, Yokohama}
\affiliation{Korea University, Seoul}
\affiliation{Kyoto University, Kyoto}
\affiliation{Kyungpook National University, Taegu}
\affiliation{Swiss Federal Institute of Technology of Lausanne, EPFL, Lausanne}
\affiliation{University of Ljubljana, Ljubljana}
\affiliation{University of Maribor, Maribor}
\affiliation{University of Melbourne, Victoria}
\affiliation{Nagoya University, Nagoya}
\affiliation{Nara Women's University, Nara}
\affiliation{National Central University, Chung-li}
\affiliation{National Kaohsiung Normal University, Kaohsiung}
\affiliation{National United University, Miao Li}
\affiliation{Department of Physics, National Taiwan University, Taipei}
\affiliation{H. Niewodniczanski Institute of Nuclear Physics, Krakow}
\affiliation{Nippon Dental University, Niigata}
\affiliation{Niigata University, Niigata}
\affiliation{Nova Gorica Polytechnic, Nova Gorica}
\affiliation{Osaka City University, Osaka}
\affiliation{Osaka University, Osaka}
\affiliation{Panjab University, Chandigarh}
\affiliation{Peking University, Beijing}
\affiliation{Princeton University, Princeton, New Jersey 08544}
\affiliation{RIKEN BNL Research Center, Upton, New York 11973}
\affiliation{Saga University, Saga}
\affiliation{University of Science and Technology of China, Hefei}
\affiliation{Seoul National University, Seoul}
\affiliation{Shinshu University, Nagano}
\affiliation{Sungkyunkwan University, Suwon}
\affiliation{University of Sydney, Sydney NSW}
\affiliation{Tata Institute of Fundamental Research, Bombay}
\affiliation{Toho University, Funabashi}
\affiliation{Tohoku Gakuin University, Tagajo}
\affiliation{Tohoku University, Sendai}
\affiliation{Department of Physics, University of Tokyo, Tokyo}
\affiliation{Tokyo Institute of Technology, Tokyo}
\affiliation{Tokyo Metropolitan University, Tokyo}
\affiliation{Tokyo University of Agriculture and Technology, Tokyo}
\affiliation{Toyama National College of Maritime Technology, Toyama}
\affiliation{University of Tsukuba, Tsukuba}
\affiliation{Utkal University, Bhubaneswer}
\affiliation{Virginia Polytechnic Institute and State University, Blacksburg, Virginia 24061}
\affiliation{Yonsei University, Seoul}
  \author{K.~Abe}\affiliation{High Energy Accelerator Research Organization (KEK), Tsukuba} 
  \author{K.~Abe}\affiliation{Tohoku Gakuin University, Tagajo} 
  \author{I.~Adachi}\affiliation{High Energy Accelerator Research Organization (KEK), Tsukuba} 
  \author{H.~Aihara}\affiliation{Department of Physics, University of Tokyo, Tokyo} 
  \author{K.~Aoki}\affiliation{Nagoya University, Nagoya} 
  \author{K.~Arinstein}\affiliation{Budker Institute of Nuclear Physics, Novosibirsk} 
  \author{Y.~Asano}\affiliation{University of Tsukuba, Tsukuba} 
  \author{T.~Aso}\affiliation{Toyama National College of Maritime Technology, Toyama} 
  \author{V.~Aulchenko}\affiliation{Budker Institute of Nuclear Physics, Novosibirsk} 
  \author{T.~Aushev}\affiliation{Institute for Theoretical and Experimental Physics, Moscow} 
% ...
\collaboration{The Belle Collaboration}

\collaboration{Belle Collaboration}
\noaffiliation

\begin{abstract}
We report an improved measurement of the branching fraction
for $\bzbar \rightarrow D_{sJ}^\ast(2317)^+ K^-$
and present evidence of the $\bzbar \rightarrow D_{sJ}(2460)^+ K^-$ decay.
These results are obtained from a data sample containing 386
million $B\bar{B}$ pairs that was collected 
near the $\Upsilon(4S)$ resonance,
with the Belle detector at the KEKB asymmetric energy $e^+ e^-$
collider.
\end{abstract}


\maketitle

\tighten

{\renewcommand{\thefootnote}{\fnsymbol{footnote}}}
\setcounter{footnote}{0}

Two narrow resonances denoted as $D_{sJ}^\ast(2317)^+$ and $D_{sJ}(2460)^+$
have been observed recently 
in $e^+ e^-$ continuum interactions~\cite{baba,cleo,bela,babb} and in
$B$ decays ~\cite{belb,babc,belc}.
The surprisingly low masses and small widths of these states 
initiated a wide theoretical discussion ~\cite{dsjaa}. Although 
the $0^+$ and $1^+$ quantum numbers have
been established for the 
$D_{sJ}^\ast(2317)^+$ and $D_{sJ}(2460)^+$ resonances \cite{fna},
respectively, the nature of these states is still unclear.

In this paper we report an updated study of the decays 
$\bzbar \to D_{sJ}^+ K^-$
with a data sample that is approximately 2.5 times larger 
than in the recently Belle published paper \cite{belc}
that first reported the $\bzbar \rightarrow D_{sJ}^\ast(2317)^+ K^-$
decay mode.
In the previous Belle study the product branching fraction
\mbox{${\cal B}(\bzbar \rightarrow D_{sJ}^\ast(2317)^+ K^-) \times$}
\mbox{${\cal B}(D_{sJ}^\ast(2317)^+ \rightarrow D_s^+ \pi^0) =$} 
\mbox{$(5.3^{+1.5}_{-1.3} \pm 0.7 \pm 1.4) \times 10^{-5}$}
was measured and an upper limit
${\cal B}(\bzbar \rightarrow D_{sJ}(2460)^+ K^-) \times
{\cal B}(D_{sJ}(2460)^+ \rightarrow D_s^+ \gamma) < 0.94 \times 10^{-5}$
was set.
These measurements~\cite{fnb} show that
$\mathcal{B}(\bzbar \to D_{sJ}^\ast(2317)^+ K^-)$ 
is of the same order of magnitude as $\mathcal{B}(\bzbar \to D_s^+ K^-)$
\cite{beld,babd} and at least a factor of two larger than
the branching fraction for $\bzbar \to D_{sJ}(2460)^+ K^-$.

The $\bzbar \to D_{s(J)}^+ K^-$ decays can be described
by a PQCD factorization $W$ exchange 
process ~\cite{kteoa,kteob} or, alternatively, by final state 
interactions ~\cite{kteoc,kteod}.
Assuming there is a four-quark component of the $D_{sJ}$ mesons,
the tree diagram with $s\bar{s}$ pair creation may also contribute \cite{belc}.
Although accurate theoretical calculations of branching fractions
are difficult for these decay modes, the experimental results
disagree with the na\"{\i}ve expectation \cite{dsji} 
that the ratio $\mathcal{B}(\bzbar \to D_{s}^+ h^-)/ 
\mathcal{B}(\bzbar \to D_{sJ}^+ h^-)$
should be similar for $h^-=\pi^-, K^-$ or $D^-$.

This analysis is based on a large data sample,
which contains 386 million $B\overline{B}$ pairs, 
collected with the Belle detector at the KEKB asymmetric-energy
$e^+e^-$ (3.5 on 8~GeV) collider~\cite{KEKB}
operating at the $\Upsilon(4S)$ resonance.
The Belle detector is a large-solid-angle magnetic
spectrometer that
consists of a silicon vertex detector (SVD),
a 50-layer central drift chamber (CDC), an array of
aerogel threshold \v{C}erenkov counters (ACC), 
a barrel-like arrangement of time-of-flight
scintillation counters (TOF), and an electromagnetic calorimeter
comprised of CsI(Tl) crystals (ECL) located inside 
a super-conducting solenoid coil that provides a 1.5~T
magnetic field.  An iron flux-return located outside of
the coil is instrumented to detect $K_L^0$ mesons and to identify
muons (KLM).  The detector
is described in detail elsewhere~\cite{Belle}.
Two inner detector configurations were used. A 2.0 cm beampipe
and a 3-layer silicon vertex detector was used for the first sample
of 152 million $B\bar{B}$ pairs, while a 1.5 cm beampipe, a 4-layer
silicon detector and a small-cell inner drift chamber were used to record  
the remaining 234 million $B\bar{B}$ pairs\cite{SVD}.  

In this analysis we applied the same selection criteria 
as in \cite{belc}, where a detailed description of
the criteria can be found. The only differences between
the two analyses arise due to the vertex detector upgrade.
According to the MC simulation,
minor differences in the signal widths and efficiencies
are expected for the two SVD subdetector configurations, leading to
respective corrections applied in the fit procedure and
efficiency calculations.

Kaon and pion mass hypotheses are assigned to the
charged tracks with momenta $p > 100\,\mev/c$~\cite{fnb} using a likelihood 
ratio ${\cal L}_{K/\pi} = {\cal L}_K/({\cal L}_K + {\cal L}_{\pi})$, obtained
by combining information from the CDC ($dE/dx$), ACC, and TOF systems.
We require ${\cal L}_{K/\pi} > 0.6$ (${\cal L}_{K/\pi} < 0.6$) 
for kaon (pion) candidates~\cite{Belle}.

ECL clusters with a photon-like shape and energies larger than 50 MeV,
that are not associated with charged tracks,
are accepted as photon candidates. Photon pairs of invariant mass within
$\pm 12\,\mev/c^2$ (\mbox{$\sim\,3\sigma$} in the $\pi^0$ mass resolution) 
of the $\pi^0$ mass are considered $\pi^0$ candidates;
the $\pi^0$ momentum is required to be larger than $100\,\mev/c$.

$K^0_S$ candidates are formed from $\pi^+\pi^-$ pairs 
with an invariant mass within $\pm 10\,\mev/c^2$ ($\sim$\,3$\sigma$)
of the nominal $K^0_S$ mass.
Invariant masses of $K^{*0} \to K^+ \pi^-$ candidates are required to
be within $\pm 50\,\mev/c^2$ of the nominal $K^{*0}$ mass; those of
$\phi \to K^+ K^-$ candidates, within $\pm 12\,\mev/c^2$ of the $\phi$ mass.
$D_s^+$ mesons are reconstructed in the $\phi \pi^+$, $\kzbarstar K^+$
and $K_S^0 K^+$ decay channels; a mass window of 
$\pm 12\,\mev/c^2$ ($\sim\,2.5\sigma$) is imposed in each case.
The $D_{sJ}$ mesons
are reconstructed in the $D_{sJ}^\ast(2317)^+ \to D_s^+ \pi^0$ and
$D_{sJ}(2460)^+ \to D_s^+ \gamma$ decay modes; within the mass difference
ranges $|M(D_s^+ \pi^0) - M(D_s^+) - 348.6| < 20\,\mev/c^2$ and
$|M(D_s^+ \gamma) - M(D_s^+) - 487.9| < 30\,\mev/c^2$.

Candidate $\bzbar \to D_{sJ}^+ K^-$ and $D_{sJ}^- \pi^+$ 
are formed and the signal is extracted using 
the energy difference $\Delta E\,=\,E^{CM}_B-E^{CM}_{\rm beam}$
and beam-constrained mass 
$M_{\rm bc}=\sqrt{(E^{CM}_{\rm beam})^2\,-\,(p^{CM}_B)^2}$;
$E^{CM}_B$ and $p^{CM}_B$ are the energy and momentum
of the $B$ candidate in the center-of-mass (CM) system
and $E^{CM}_{\rm beam}$ is the CM beam energy.
Only events within the intervals 
$M_{\rm bc} > 5.2\,\gev/c^2$ and $|\Delta E|\,<0.2\,\gev$ are used
in this analysis.
The $B$ meson signal region is defined by
$|\Delta E|\,<0.04\,\gev$ and 
$5.272\,\gev/c^2\, < M_{\rm bc} < 5.288\,\gev/c^2$.

Combinatorial background for channels involving the 
$D_{sJ}(2460)^+$ was further suppressed by requiring
\mbox{$\cos \theta_{D_{s}\gamma} < 0.7$}.
The helicity angle $\theta_{D_{s}\gamma}$ is
defined as the angle between 
the direction opposite the $B$ momentum and 
the $D_s^+$ momentum in the $D_s^+ \gamma$ rest frame.
This requirement rejects 49$\%$ of background 
events and only 6$\%$ of signal events, assuming \mbox{$J^P = 1^+$} 
for the $D_{sJ}(2460)^+$. The uncertainty due to
this assumption is included in the systematic error.

For events with two or more $B$ candidates,
the $D_s^+$ and $\pi^0$ candidates
with invariant masses closest to their nominal values
and the $B$ daughter $K^+$ or $\pi^-$ candidate 
with the best ${\cal L}_{K/\pi}$ value are chosen.
No multiple entries are found in the data.

We exploit the event topology to separate $B\bbar$ events (spherical)
from the continuum background (jetlike).
The ratio of the second and zeroth Fox-Wolfram moments~\cite{SFW} 
of all particles in the event is required to be less than 0.5.
For such events, we form a Fisher discriminant
from six modified Fox-Wolfram moments.
A signal (background) likelihood ${\cal L}_S$ (${\cal L}_{BG}$) is obtained
using signal MC (sideband) data from the product of probability density
functions for the Fisher discriminant and $\cos\theta_B$, where
$\theta_B$ is the $B$ flight direction in the CM system 
with respect to the $z$ axis.
We require  
${\cal R} = {\cal L}_S/({\cal L}_S + {\cal L}_{BG}) > 0.4$ for 
$D_s^+ \to \kzbarstar K^+$ and ${\cal R} > 0.25$ for the other
$D_s^+$ decay modes, which have lower backgrounds.

\begin{figure}[htb]
\includegraphics[width=0.41\textwidth]{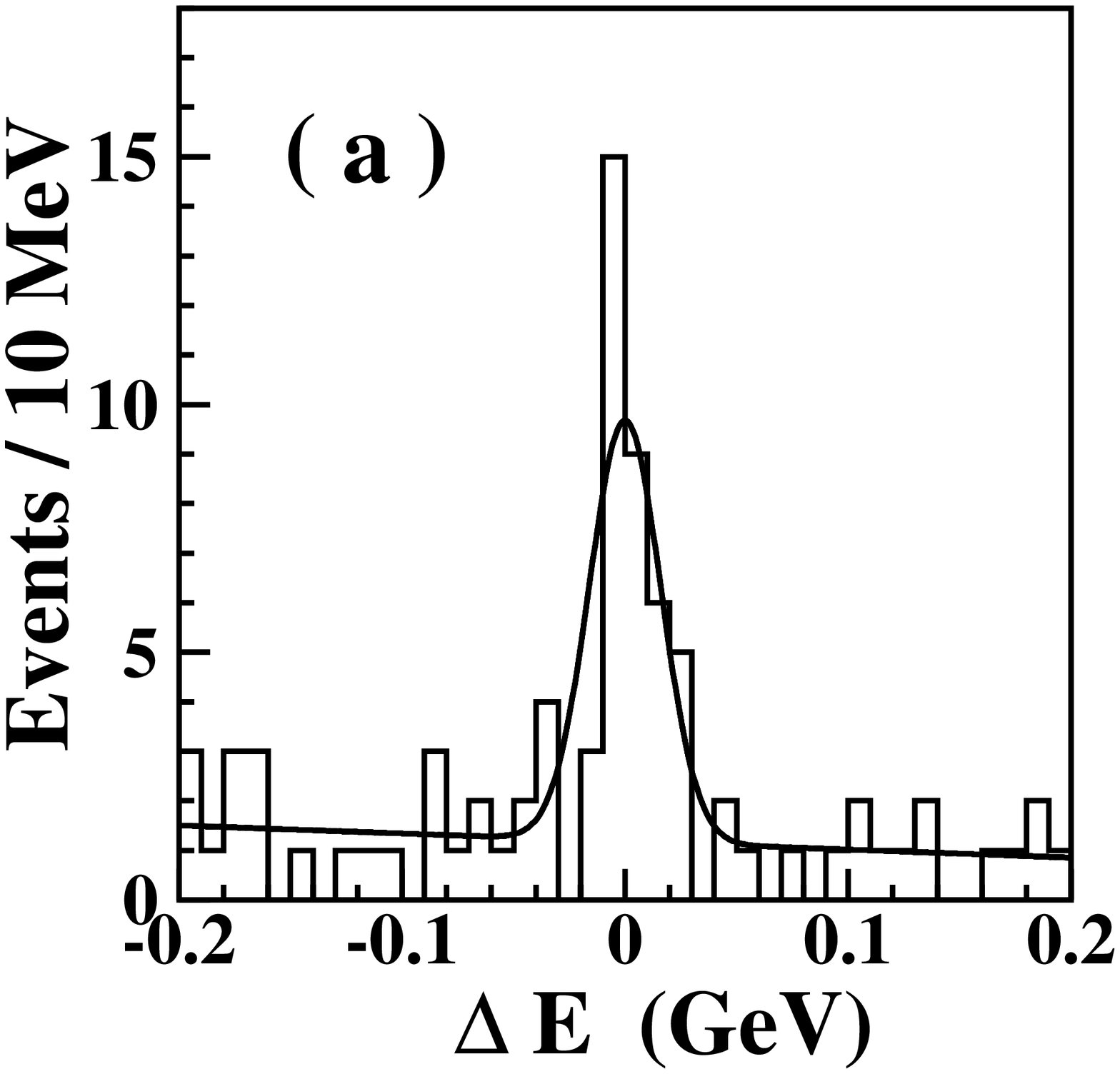}\includegraphics[width=0.41\textwidth]{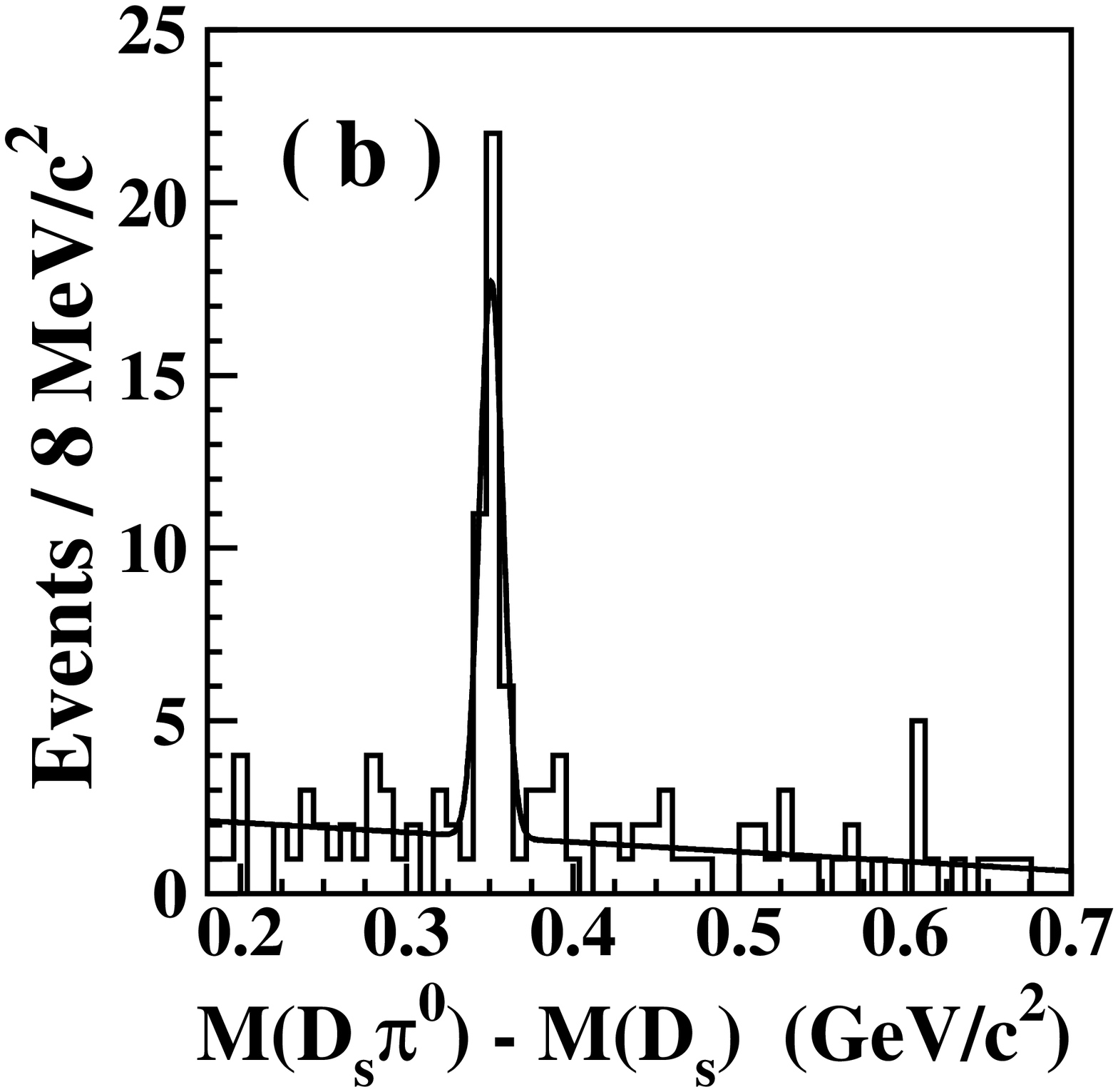}
\includegraphics[width=0.41\textwidth]{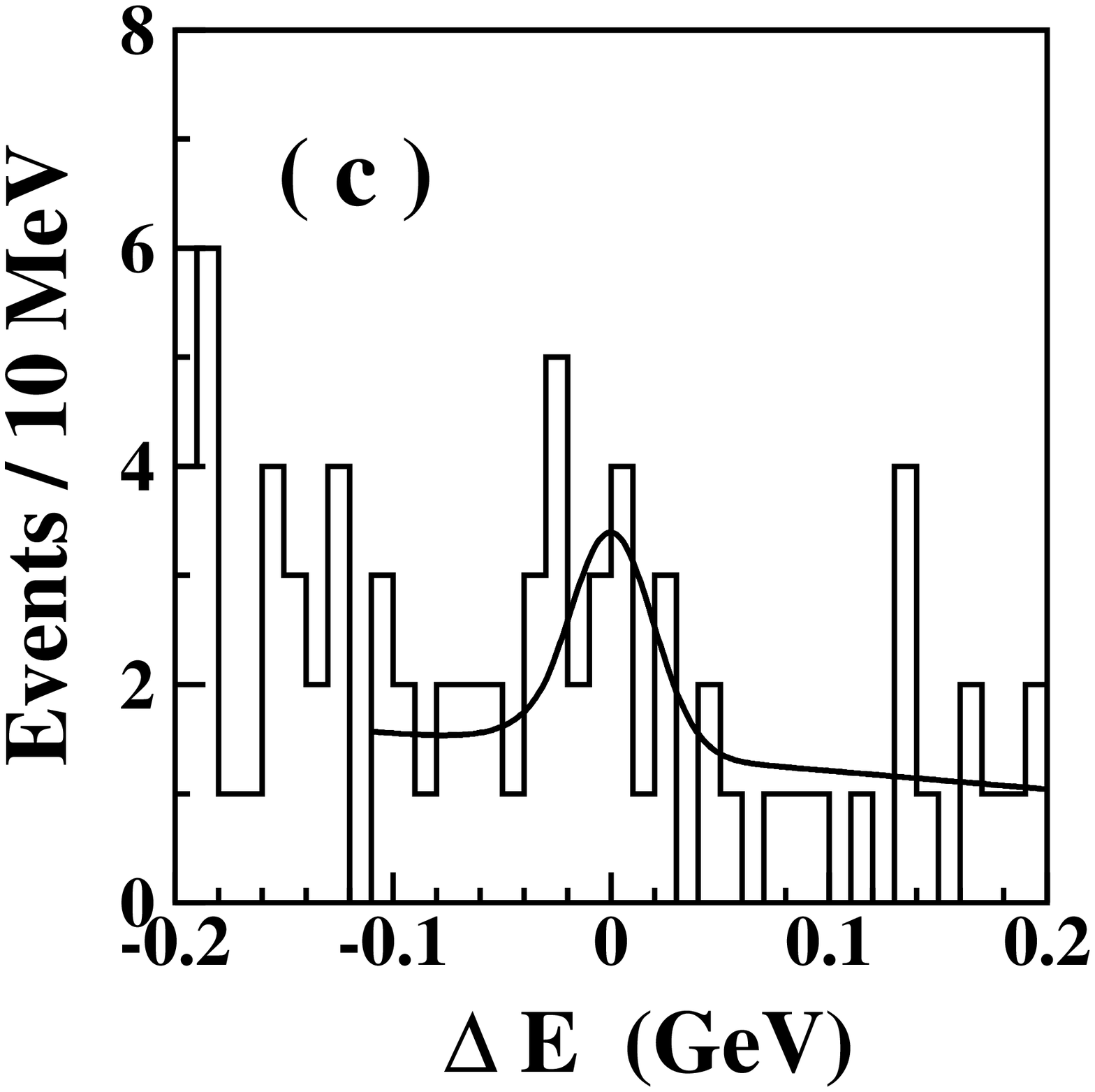}\includegraphics[width=0.41\textwidth]{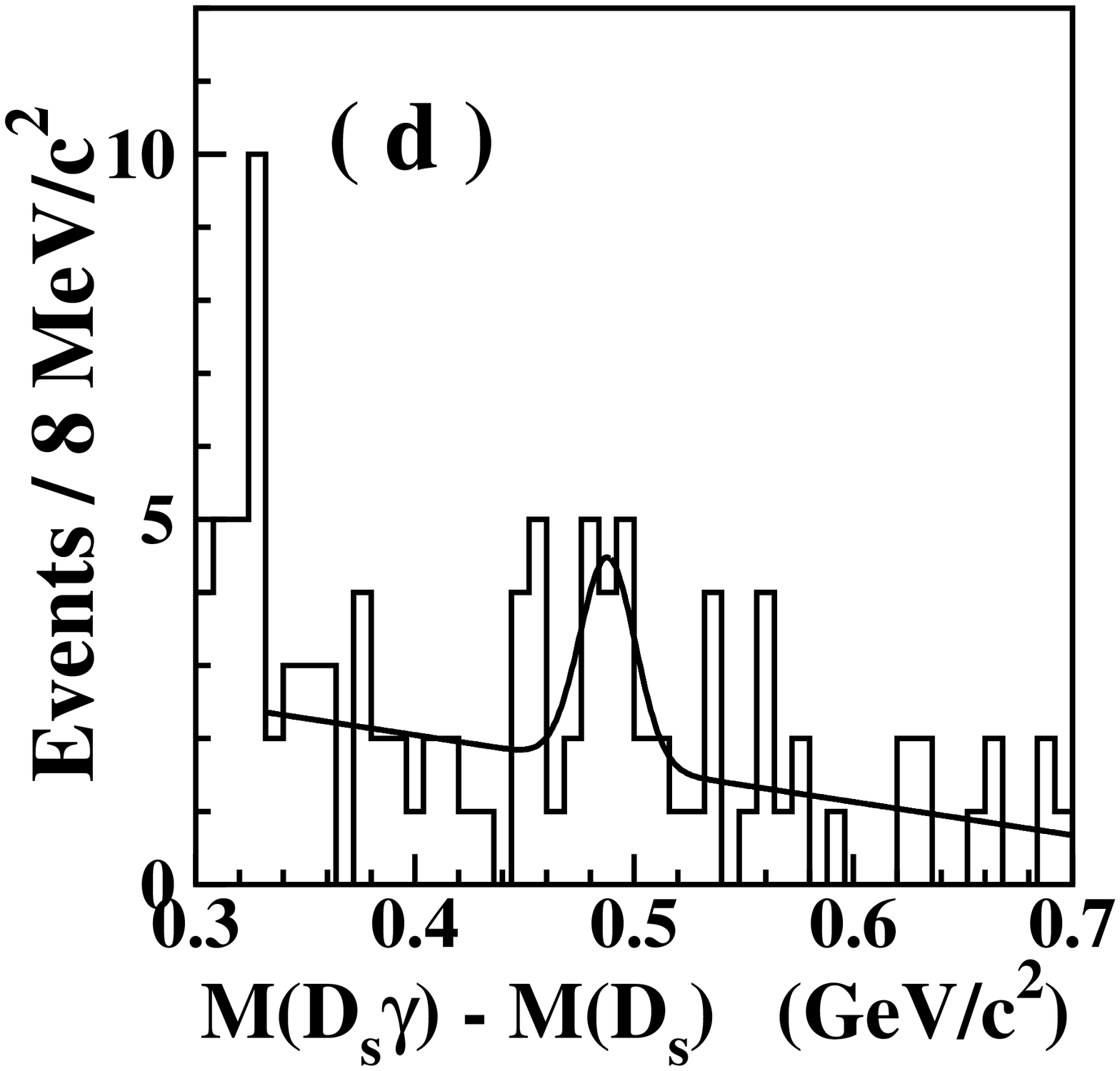}
\vspace{-0.3cm}
\caption{$\Delta E$ (a) and $\Delta M(D_{sJ})$ (b) distributions
for the $\bzbar \to D_{sJ}^\ast(2317)^+ K^-$ decay,
and $\Delta E$ (c) and $\Delta M(D_{sJ})$ (d) distributions
for the $\bzbar \to D_{sJ}(2460)^+ K^-$ decay.}
\label{figall}
\end{figure}

The $\Delta E$ and $\Delta M(D_{sJ})$ distributions
for the $D_{sJ}^+ K^-$ combinations are shown in Fig.~\ref{figall}
for the range $5.272\,\gev/c^2\, < M_{\rm bc} < 5.288\,\gev/c^2$.
To obtain the $\Delta M(D_{sJ})$
distributions we relax the $\Delta M(D_{sJ})$ requirements and
apply the tight selection on $\Delta E$.
The $\Delta E$ distributions are modelled using 
a linear background function and a 
Gaussian signal shape (the Crystal Ball shape
function \cite{cb} is used for the $D_{sJ}(2460)^+$)
with zero mean and    
a fixed width determined from MC data. 
The $\Delta M(D_{sJ})$ distributions are described by
the sum of a signal Gaussian and a linear background.
The widths of the Gaussians are fixed from MC while the
peak positions are allowed to float.
A strong $\bzbar \to D_{sJ}^\ast(2317)^+ K^-$ signal is observed
and evidence of the $\bzbar \to D_{sJ}(2460)^+ K^-$ signal
is also seen. The Gaussian peak positions
obtained from the fits correspond to the $D_{sJ}$ mass values
of $2319.2 \pm 1.3\,$MeV/c$^2$ and $2456.2 \pm 6.5\,$MeV/c$^2$
for the $D_{sJ}^\ast(2317)^+$ and $D_{sJ}(2460)^+$, respectively.
These values are in a good agreement with the most recent BaBar
measurements \cite{babam} of $D_{sJ}$ masses in the continuum,
$2318.9 \pm 0.3 \pm 0.9\,$MeV/c$^2$ 
and $2459.4 \pm 0.3 \pm 1.0\,$MeV/c$^2$.

Signal yields, efficiencies, branching fractions and significances
for the studied decay channels are shown
in Table 1.
The signal yields are obtained from the fits of histograms
shown in Fig.~\ref{figall}, where the three $D_s$ decay channels are combined.
The $\bzbar \rightarrow D_{sJ}^+ K^-$ branching fractions and
significances are obtained
using a simultaneous fit to the $\Delta M(D_{sJ})$ distributions 
for the three $D_s^+$ decay channels, with independent background descriptions,
but common values for the signal width (fixed from MC)
and peak position (allowed to float).
The branching fractions obtained in the individual decay modes agree
within the statistical errors. 
The three error terms are the statistical uncertainty, the total systematic
error, and the uncertainty due to $D_s^+$ branching fractions;
this last term is dominated by the $\sim25\%$ uncertainty
in ${\cal B}(D_s^+ \to \phi \pi^+)$ \cite{pdg}.
The last two systematic terms are combined for the 
$\bzbar \to D_{sJ}(2460)^+ K^-$ decay.
The significance is defined as
$\sqrt{-2\,\ln({\cal L}_0/{\cal L}_{\rm max})}$,
where ${\cal L}_{\rm max}$ and ${\cal L}_0$ are likelihoods
for the best fit and zero signal yields, respectively.
The significance is corrected for systematics due to
the peaking background, which is estimated using $\Delta E$ and
$M_{\rm bc}$ sidebands.
Efficiencies include all intermediate resonance 
branching fractions \cite{pdg} 
and were obtained from MC simulation, 
assuming $J^P = 0^+$ for the $D_{sJ}^\ast(2317)$
and $J^P = 1^+$ for the $D_{sJ}(2460)$.
We assume equal production of neutral and charged $B$ mesons.

\renewcommand{\arraystretch}{1.2}
\begin{table}[htb]
\caption{Signal yields, efficiencies, product branching fractions,
and significances for the $\bzbar \to D_{sJ}^+ K^-$ processes. 
The first error is the statistical uncertainty, the second is the 
systematic uncertainty, and the third error is the uncertainty due
to $D_s^+$ decay branching fractions.
Product branching fractions are obtained from 
simultaneous $\Delta M(D_{sJ})$ fits of three $D_s$ decay modes
as described in the text.}
\label{bfr}
\begin{tabular}
{@{\hspace{0.2cm}}l@{\hspace{0.2cm}}||@{\hspace{0.2cm}}c@{\hspace{0.2cm}}|@{\hspace{0.2cm}}c@{\hspace{0.2cm}}|@{\hspace{0.2cm}}c@{\hspace{0.2cm}}|@{\hspace{0.2cm}}c@{\hspace{0.2cm}}|@{\hspace{0.2cm}}c@{\hspace{0.2cm}}}
\hline \hline
Decay mode      & Yield & Yield & Efficiency & Product $\mathcal{B}(\bzbar\to D_{sJ}^+ K^-) \times$  & Signif.  \\
                & $\Delta M(D_{sJ})$ & $\Delta E$  & $(10^{-4})$ & $\mathcal{B}(D_{sJ} \to D_s \pi^0 (\gamma))\; (10^{-5})$ & $\sigma$   \\ \hline 
$D_{sJ}^\ast(2317)^+K^-$ & $35.3 \pm 6.4$ & $34.1 \pm 6.6$  & $21.9 \pm 0.6$ & $4.4 \pm 0.8 \pm 0.6 \pm 1.1$        & 9.2   
\\
$D_{sJ}(2460)^+ K^-$    & $11.2 \pm 5.4$ & $10.2 \pm 5.4$ & 
$59.5 \pm 1.4$     & $0.53 \pm 0.20 ^{+0.16}_{-0.15}$ & 3.1 \\
 & & & & $<0.86\,(90\%\,\text{C.L.})$ &  \\ 
\hline \hline
\end{tabular}
\end{table}

The PDG value of ${\cal B}(D_s^+ \rightarrow \phi \pi^+) = (3.6 \pm 0.9)\%$
\cite{pdg} with a 25$\%$ uncertainty is used to obtain the branching 
fractions listed in Table 1.
BaBar has recently determined the branching fraction
${\cal B}(D_s^+ \rightarrow \phi \pi^+) = (4.81 \pm 0.52 \pm 0.38)\%$ 
\cite{babah}, which has a smaller uncertainty of 13$\%$.
If we use this BaBar value 
the product branching fractions become
\mbox{${\cal B}(\bzbar \rightarrow D_{sJ}^\ast(2317)^+ K^-) \times
{\cal B}(D_{sJ}^\ast(2317)^+ \rightarrow D_s^+ \pi^0) =
(3.3 \pm 0.6 \pm 0.7) \times 10^{-5}$} and
\mbox{${\cal B}(\bzbar \rightarrow D_{sJ}(2460)^+ K^-) \times
{\cal B}(D_{sJ}(2460)^+ \rightarrow D_s^+ \gamma) =
(0.40 \pm 0.15 ^{+0.12}_{-0.11}) \times 10^{-5}$}. 
The major sources contributing to the systematic error are shown in Table 2.
More details about the systematic uncertainties 
can be found in \cite{belc}.

\begin{table}[htb]
\caption{Systematic uncertainties in the 
$\bzbar \rightarrow D_{sJ}^+ K^-$ branching fraction measurements.}
\label{sys1}
\begin{tabular}
{@{\hspace{0.5cm}}l@{\hspace{0.5cm}}||@{\hspace{0.5cm}}c@{\hspace{0.5cm}}|@{\hspace{0.5cm}}c@{\hspace{0.5cm}}}
\hline \hline
       & \multicolumn{2}{c}{Systematic error (\%)} \\ \cline{2-3}
Source & $D_{sJ}^\ast(2317)^+ K^-$ & $D_{sJ}(2460)^+ K^-$ \\
\hline
Track reconstruction  & $\pm 1 \times N_{tracks}$ & $\pm 1 \times N_{tracks}$ \\
Charged particle identification & $\pm 2 \times N_{particles}$ & $\pm 2 \times N_{particles}$ \\
Photon and $\pi^0$ reconstruction & $\pm 5$ & $\pm 2$ \\
$K_S^0$ reconstruction & $\pm 3$ & $\pm 3$ \\
$\Delta E$ and likelihood ratio shapes & $\pm 4$ & $\pm 4$ \\
Helicity angular distribution assumption & $\pm 4$ & $^{+9}_{-0}$  \\
Background subtraction & $\pm 6$ & $\pm 5$ \\
Fitting procedure & $\pm 3$ & $\pm 5$ \\
MC statistics & $\pm 2$ & $\pm 2$ \\
Number of $B\bar{B}$ pairs & $\pm 1.5$ & $\pm 1.5$ \\
\hline
Total & $\pm 14$ & $^{+16}_{-13}$ \\
\hline \hline
\end{tabular}
\end{table}

In conclusion, improved measurements of
$\bzbar \to D_{sJ}^+ K^-$ decay modes have been performed 
using a data sample approximately 2.5 times larger. 
Good agreement with the previous measurement is obtained \cite{belc}.
The value of $\mathcal{B}(\bzbar \to D_{sJ}^\ast(2317)^+ K^-)$ 
is of the same order of magnitude as $\mathcal{B}(\bzbar \to D_s^+ K^-)$
and significantly larger than
the $\bzbar \to D_{sJ}(2460)^+ K^-$ branching fraction.

We thank the KEKB group for the excellent operation of the
accelerator, the KEK cryogenics group for the efficient
operation of the solenoid, and the KEK computer group and
the National Institute of Informatics for valuable computing
and Super-SINET network support. We acknowledge support from
the Ministry of Education, Culture, Sports, Science, and
Technology of Japan and the Japan Society for the Promotion
of Science; the Australian Research Council and the
Australian Department of Education, Science and Training;
the National Science Foundation of China under contract
No.~10175071; the Department of Science and Technology of
India; the BK21 program of the Ministry of Education of
Korea and the CHEP SRC program of the Korea Science and
Engineering Foundation; the Polish State Committee for
Scientific Research under contract No.~2P03B 01324; the
Ministry of Science and Technology of the Russian
Federation; the Ministry of Higher Education, 
Science and Technology of the Republic of Slovenia;  
the Swiss National Science Foundation; the National Science Council and
the Ministry of Education of Taiwan; and the U.S.\
Department of Energy.


\begin{thebibliography}{99}
\bibitem{fna}
Charge conjugate modes are implicitly included everywhere.
\bibitem{fnb}
Recent measurements of $D_{sJ}^\ast(2317)^+$ and $D_{sJ}(2460)^+$ decay
branching fractions indicate that the
$D_{sJ}^\ast(2317)^+ \to D_s^+ \pi^0$ channel is dominant and the 
$D_{sJ}(2460)^+ \to D_s^+ \gamma$ branching fraction is around 30$\%$.
\end{thebibliography}
\end{document}